\documentclass[%
 reprint,
amsmath,amssymb,
prb,
]{revtex4-2}

\usepackage{graphicx}
\usepackage{dcolumn}
\usepackage{bm}

\usepackage{hyperref} 
\hypersetup{
	colorlinks=true,
	linkcolor=blue,
	citecolor=blue,
	filecolor=blue,      
	urlcolor=blue,
}
\usepackage{hypcap}

\usepackage[normalem]{ulem} 
\usepackage[dvipsnames]{xcolor} 

\newcommand{\comentar}[1]{}

\graphicspath{{Images/}}

\setcitestyle{numbers,square}

\newcommand{\tint}[1]{\int_{-\infty}^{\infty} {#1}\,dt}
\newcommand{\fou}[1]{\tint{#1 e^{i\omega t}}}
\newcommand{\ifou}[2]{\int_{-\infty}^{\infty} {#1} e^{-i #2 t}\,\frac{d #2}{2\pi}}

\DeclareMathAlphabet\mathbfcal{OMS}{cmsy}{b}{n}

\newcommand{\tensb}[1]{\tensor{#1}}
\newcommand{\vv}[1]{\vec{#1}}

\newcommand{\w}[0]{\omega }

\usepackage{empheq}
\usepackage[most]{tcolorbox}
\newtcbox{\mymath}[1][]{%
	nobeforeafter, math upper, tcbox raise base,
	enhanced, colframe=orange!90!white,
	colback=orange!2!white, boxrule=1.5pt,
	#1}
\usepackage{tcolorbox}
\newtcolorbox{mybox}[1]{ colframe=orange!94!black, colback=orange!2!white,fonttitle=\bfseries,title=#1}
\newtcolorbox{mybox2}[1]{ colframe=blue!94!black, colback=blue!2!white,fonttitle=\bfseries,title=#1}

\begin{document}

\title{Theory and simulations of the angular momentum transfer from swift electrons to spherical nanoparticles in STEM}




\author{Jos\'e \'Angel Castellanos-Reyes}
\altaffiliation{Currently at Department of Physics and Astronomy, Uppsala University, Box 516, 75121 Uppsala, Sweden}
\email{angel.castellanos.research@gmail.com}
\affiliation{Departamento de F\'isica, Facultad de Ciencias, Universidad Nacional Aut\'onoma de M\'exico, Ciudad Universitaria, Av. Universidad $\# 3000$, Mexico City, 04510, Mexico.}

\author{Jes\'us Castrej\'on-Figueroa}%
\affiliation{Departamento de F\'isica, Facultad de Ciencias, Universidad Nacional Aut\'onoma de M\'exico, Ciudad Universitaria, Av. Universidad $\# 3000$, Mexico City, 04510, Mexico.}
\author{Alejandro Reyes-Coronado}
\email{coronado@ciencias.unam.mx}
\affiliation{Departamento de F\'isica, Facultad de Ciencias, Universidad Nacional Aut\'onoma de M\'exico, Ciudad Universitaria, Av. Universidad $\# 3000$, Mexico City, 04510, Mexico.}

\date{\today}


\begin{abstract}
Electron beams in scanning transmission electron microscopy (STEM) exert forces and torques on study samples, with magnitudes that allow the controlled manipulation of nanoparticles (a technique called electron tweezers). Related theoretical research has mostly focused on the study of forces and linear momentum transfers from swift electrons (like those used in STEM) to nanoparticles. However, theoretical research on the rotational aspects of the interaction would benefit not only the development of electron tweezers, but also other fields within electron microscopy such as electron vortices studies. Starting from a classical-electrodynamics description, we present a theoretical model, alongside an efficient numerical methodology, to calculate the angular momentum transfer from a STEM swift electron to a spherical nanoparticle. We show simulations of angular momentum transfers to aluminum, gold, and bismuth nanoparticles of different sizes. We found that the transferred angular momentum is always perpendicular to the system's plane of symmetry, displaying a constant direction for all the cases considered. In the simulations, the angular momentum transfer increased with the radius of the nanoparticle, but decreased as the speed of the electron or the impact parameter increased. Also, the electric contribution to the angular momentum transfer dominated over the magnetic one, being comparable only for high electron’s speeds (greater than 90\% of the speed of light). Moreover, for nanoparticles with 1 nm radius of the studied materials, it was found that the small-particle approximation (in which the nanoparticles are modeled as electric point dipoles) is valid and accurate to compute the angular momentum transfer as long as the impact parameter is greater than four times the nanoparticle’s radius, and that the electron’s speed exceeds 50\% of the speed of light. We believe that these findings contribute to the understanding of rotational aspects present in STEM experiments, and might be useful for further developments in electron tweezers and other electron microscopy related techniques.
\end{abstract}

\maketitle

%
%

\section{Introduction}

Transmission electron microscopy (TEM) has played a decisive role in the study and characterization of micro and nanostructures \cite{Krivanek1,Krivanek2,deabajo2021optical,maclaren2020detectors,GarciadeAbajo-1}.
Interestingly, the electron beams used in TEM have the potential to become effective tools for the controlled manipulation of nanostructures  \cite{OLESHKO2013203}. The forces and torques exerted on nanoparticles (NPs) by  TEM electron beams have been exploited to successfully control their movement in a technique that has been called \textit{electron tweezers}, in analogy to optical tweezers \cite{OLESHKO2013203,Oleshko,zheng2012electron,verbeeck2013,Vesseur,Batson,Batson2,Batson3}. Particular interest has been placed on the study of the interaction of NPs with TEM electron beams in the scanning mode (STEM) \cite{Batson,Batson01,GarciadeAbajo0,PRBCoronado,Vesseur,Batson2,Batson3,Lagos,Lagos2,castellanos2019,castellanos2021angular,castrejon2021time,PRBcastrejon2}, due to its high spatial  $\left(\text{reaching} < 0.1 \text{ nm}\right)$ and spectral (up to $< 5$ meV) resolutions  \cite{Krivanek1,Krivanek2,Krivanek2019}, which are constantly improving. 

Most of the theoretical work related to electron tweezers has focused on the study of the force and linear momentum that a swift electron (from a STEM electron beam) transfers to NPs. Even though there are some works on the torque and the angular momentum transfer to small NPs \cite{GarciadeAbajo0, GarciadeAbajo-1, castellanos2021angular}, a general study is still pending. Moreover, knowledge of the angular dynamics that electron beams induce in NPs would also benefit other fields within electron microscopy, such as electron vortices \cite{Lloyd}. 

In this work, we present a theoretical study and numerical simulations of the angular momentum that a swift electron from a STEM electron beam (without vorticity) transfers to a spherical NP. In particular, we discuss and implement an efficient numerical methodology, allowing high-precision computations of the angular momentum transfer. We show results for aluminum, gold, and bismuth NPs of different sizes within the nanoscale. Additionally, we test the applicability of the so-called \textit{small-particle approximation} for the angular momentum transfer (in which the NPs are modeled as electric point dipoles)  \cite{castellanos2021angular}, studying NPs with radii 1 nm of the studied materials.

%
%

\section{Model for the angular momentum transfer from STEM electron beam to spherical nanoparticles}
\label{Theoreticalmodel} 

Currently, STEM electron beams can be focused on spots as small as 0.05 nm, with electric currents on the order of pA and energies ranging typically from 100 keV to 400 keV, reaching even 1 MeV \cite{Krivanek1,Krivanek2,VERBEECK2005324,OLESHKO2013203}. These electron beams consist, effectively, of trains of swift electrons with relativistic speeds, between $55\%$ and $94\%$ the speed of light, each of which is emitted approximately every $10^{-8}$ s. 

In contrast, the lifetime of electronic excitations in NPs is typically on the order of $10^{-14}$ s \cite{Ashcroft,quijada2010lifetime}. Therefore, in STEM experiments, the NPs interact with the electron beam, practically one swift electron at a time.
 	
Moreover, the electron beams in STEM studies remain practically straight, with negligible deflection (on the order of milliradians) as long as they do not directly impinge on the samples \cite{Rivacoba1,deabajo2021optical,krehl2018spectral}. In addition, STEM-electron-beam energy losses (ranging from 0 to hundreds of eV) are typically much smaller than routine initial beam energies (hundreds of keV) \cite{egerton2011electron,GarciadeAbajo-1}. Hence, changes in kinetic energy of STEM swift electrons and, consequently, changes in their speed, are negligible. Thus, for many studies \cite{GarciadeAbajo0,Batson,PRBCoronado,Batson2,Batson3,Lagos,Lagos2,castellanos2019,castrejon2021time, castellanos2021angular}, and in particular for this work, it is sufficient to study the interaction between a NP and a single swift electron traveling in a straight aloof trajectory with a constant relativistic speed.

Quantum effects can be important in the interaction between a swift electron and a NP. However, there are situations in which a classical-electrodynamics picture is sufficient to obtain accurate results, as is the case for electron energy loss spectroscopy and cathodoluminescence \cite{deabajo2021optical}. This has motivated the study of the transfer of linear and angular momentum from a swift electron to a NP using classical models \cite{GarciadeAbajo0,Batson,PRBCoronado,Batson2,Batson3,Lagos,Lagos2,castellanos2019,castrejon2021time, castellanos2021angular}. 

Although an electron could be considered to have a finite size due to vacuum fluctuations, its spatial extension is estimated to be on the order of its Compton wavelength \cite{milonni2013quantum}, whose value is $\lambda_\text{C}=2.426 \times 10^{-3}$ nm \cite{CODATAComptonwavelenghtelectron}. Moreover, a STEM swift electron (traveling with relativistic velocity) has a de Broglie wavelength $\lambda_\text{B}$ on the order of  $10^{-2}$ nm \cite{GarciadeAbajo-1}. Therefore, in this work we consider distances between the NP and the swift electron trajectory in the order of nanometers, always greater than both $\lambda_\text{C}$ and $\lambda_\text{B}$. Hence, it will be assumed that the electron is a classical electrically charged point particle.

In the following Subsection, we develop mathematical expressions for the angular momentum that a swift electron transfers to a spherical NP from a classical-electrodynamics description of the interaction. The NP is characterized by a scalar frequency-dependent dielectric function $\varepsilon\left(\omega\right)$. Unless otherwise indicated, in this work we will use Gaussian atomic units (in which the electron's rest mass and electric charge, as well as the reduced Planck's constant $\hbar$ are set to 1 \cite{Drake}). However, in the presentation of our results, SI units will be used.

\subsection{Angular momentum transfer from a swift electron to a spherical nanoparticle}

\begin{figure}[h]	
	\centerline{\includegraphics [scale=0.3]{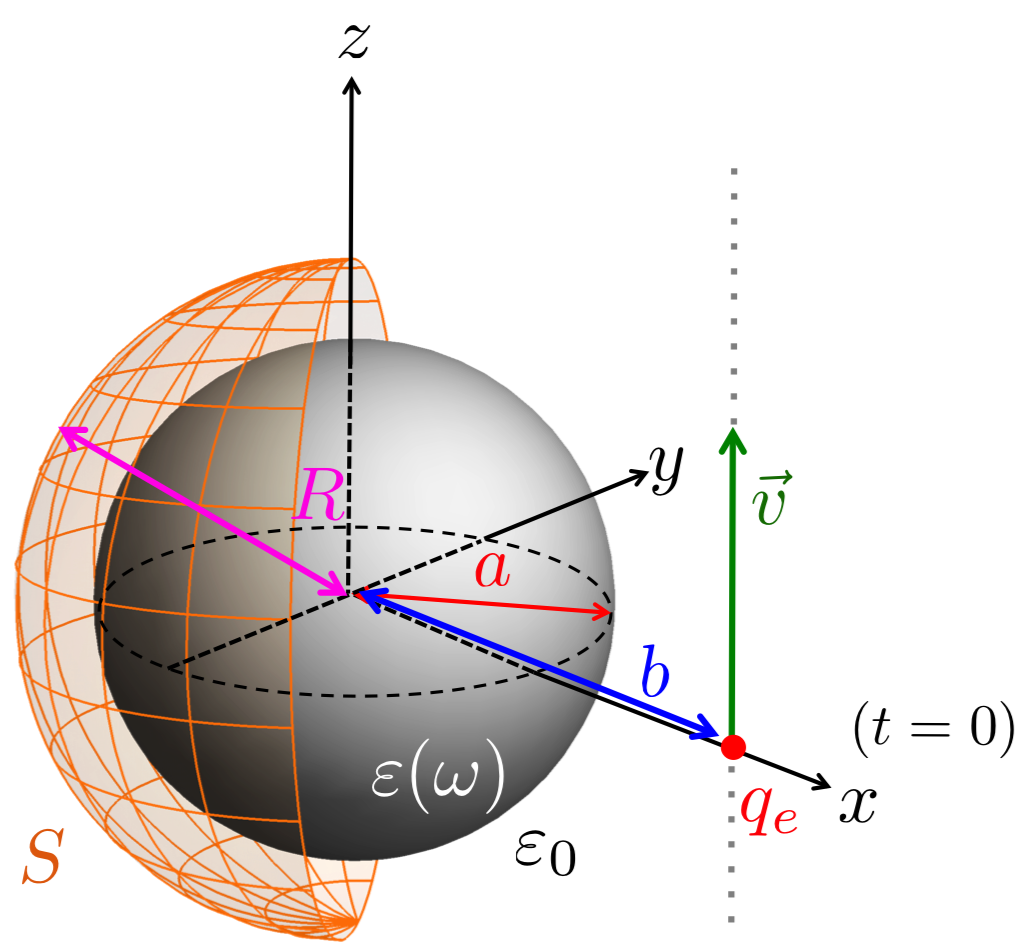}}
	\caption[Scheme of the system]{An electron, with charge $q_e$, travels with constant velocity $\vec{v}$, parallel to $z$ axis, at distance $b$ from the center of a spherical nanoparticle, of radius $a$, embedded in vacuum. The electromagnetic response of the nanoparticle is given by its dielectric function $\varepsilon(\omega)$. Time $t=0$ corresponds to the instant in which the electron is at point $(b, 0,0)$. A hemisphere of integration sphere $S$ (with radius $a <R<b$), used for the application of Eq. (\ref{angularmomentumconservation3}), is shown.}
	\label{system}
\end{figure}

Following the previous discussion, we consider a single swift electron (a point particle with charge $q_e$) traveling with constant velocity $\vec{v}$ in a straight line, at a distance $b$ (the impact parameter) from the center of a spherical nanoparticle of radius $a<b$, embedded in vacuum, as illustrated in Fig. \ref{system}. We set time $t=0$ when the electron is at its closest position to the NP and consider it travels from $t=-\infty$ to $t=\infty$, parallel to $z$ axis. Furthermore, it is assumed that the NP is electrically neutral and that its electromagnetic response is given by a frequency-dependent complex scalar dielectric function $\varepsilon(\omega)$. We consider a Cartesian coordinate system with its origin at the center of the NP, oriented so that the electron's path is parallel to the $z$-axis, intersecting the $x$-axis.
%
%

The angular momentum that the swift electron transfers to the spherical nanoparticle, with respect to the NP's center, can be obtained from the angular momentum conservation law in electrodynamics. Explicitly, for a concentric spherical surface $S$ of radius $R$ ($a<R<b$) enclosing a volume $V$ (see Fig. \ref{system}), it holds that \cite{Good}
%
\begin{equation}
\oint_S \tensb{M} \cdot d\vv{a}
=
\frac{d}{dt}\left( \vv{L}^\text{mech} + \vv{L}^\text{em}  \right),
\label{angularmomentumconservation3}
\end{equation}
%
where $\vv{L}^\text{mech}$ and $\vv{L}^\text{em}$ are the mechanical and electromagnetic angular momenta in $V$, respectively. In particular, $\vv{L}^\text{em}$ can be expressed in terms of the total electromagnetic fields as \cite{Good}
%
\begin{equation}
\vv{L}^\text{em}(t)
=
\int_V 
\vec{r}\times
\left(
\frac{\vv{E} \times \vv{B}}{4\pi c}
\right)
dV,
\label{Lem}
\end{equation}
%
where $\vec{r}$ is the position vector, $c$ the speed of light, $\vv{E}=\vv{E}\left(\vv{r},t\right)$ the electric field, and $\vv{B}=\vv{B}\left(\vv{r},t\right)$ the magnetic field. In Eq. (\ref{angularmomentumconservation3}), $\tensb{M}$ is the tensor whose $kj$ entry is
%
\begin{equation}
M_{kj}
=
\epsilon_{kli}r_lT_{ij},
\label{tensorM}
\end{equation}
%
(we use Einstein summation convention in Eq. (\ref{tensorM}) and throughout the work unless otherwise stated) being $\epsilon_{kli}$ Levi-Civita symbol, $r_l$ the $l$ component of $\vec{r}$, and $T_{ij}$ the $ij$ entry of Maxwell stress tensor, given by  \cite{Good}
%
\begin{equation}
T_{ij}
=
\frac{1}{4\pi}
\left[
E_iE_j+B_iB_j
-
\frac{\delta_{ij}}{2}\left(E^2+B^2\right)
\right]
,
\label{maxwellstresstensor}
\end{equation}
%
%
in which $E^2=\vv{E}\cdot\vv{E}$, $B^2=\vv{B}\cdot\vv{B}$, and $\delta_{ij}$ is Kronecker delta.

The total angular momentum transferred to the NP is obtained by integrating Eq. (\ref{angularmomentumconservation3}) in all the interaction time:
%
\begin{align}
&\int_{-\infty}^{\infty}\!\oint_S \!\tensb{M}\! \cdot \!d\vv{a}\,dt
=
\int_{-\infty}^{\infty}\!\frac{d}{dt}\!\left( \vv{L}^\text{mech} \!+\! \vv{L}^\text{em}  \right)\!dt
\noindent\nonumber\\
&\!
=
\!\!
\left[\vv{L}^\text{mech}(\infty) \!-\! \vv{L}^\text{mech}(\!-\infty)\right]
\!\!
+
\!\!
\left[\vv{L}^\text{em}(\infty) \!-\! \vv{L}^\text{em}(\!-\infty)\right]
\noindent\nonumber\\
&
=
\Delta \vv{L}^\text{mech} + \Delta \vv{L}^\text{em}.
\label{angularmomentum1}
\end{align}
%

There are two electromagnetic (EM) fields involved in this problem: those produced by the swift electron, called hereafter \textit{external}, whose analytical expressions (in frequency space) can be found in Ref. \cite{maciel2019electromagnetic}, and the EM fields \textit{scattered} by the NP, whose analytical expressions (in frequency space) can be found in Ref. \cite{GarciadeAbajo}. Both EM fields have been presented concisely in Appendix A of Ref. \cite{castrejon2021time}. In particular, it is worth mentioning that the scattered EM fields can be expressed as a spherical multipole expansion in the form \cite{castrejon2021time}
%
\begin{align}
	\vec{E}^\text{scat}
	&=
	\sum_{\ell=1}^{\infty} 
	\sum_{m=-\ell}^{\ell}
	\vec{E}^\text{scat}_{\ell,m},
	\label{scatteredElectricField}\\
	\vec{B}^\text{scat}
	&=
	\sum_{\ell=1}^{\infty} 
	\sum_{m=-\ell}^{\ell}
	\vec{B}^\text{scat}_{\ell,m},
	\label{scatteredMagneticField}
\end{align}
%
%
where $\ell=1$ is the dipole term, $\ell=2$ is the quadrupole term, and so on. Notably, the existence of analytical expressions for these EM fields simplifies the task at hand, since it not necessary to numerically solve Maxwell equations.

At $t = - \infty $, when the swift electron is infinitely far from the NP, the external EM fields are zero within the NP, so there are no EM fields scattered at $t=-\infty$. Therefore, from Eq. (\ref{Lem}) it follows that $ \vv{L}^\text{em} (t = -\infty)=\vv{0}$ in Eq. (\ref{angularmomentum1}).  At $ t=\infty$, the swift electron is again infinitely far from the NP, so that the external EM fields within the NP are again zero. However, during its travel, the electron induces charge and current densities within the NP, producing scattered EM fields. Nevertheless, due to dissipative effects of the NP, represented by the imaginary part of its dielectric function, all induced current and charge densities disappear at $t=\infty$. Thus, the EM fields scattered in $V$ are zero at $t =\infty$ and, consequently, $\vv{L}^\text{em}(t =\infty)=\vv{0} $. Therefore, $ \Delta \vv{L}^\text{em}=\vv{0}$ in Eq. (\ref{angularmomentum1}). 

Substituting Eqs. (\ref{tensorM}) and (\ref{maxwellstresstensor}) in Eq. (\ref{angularmomentum1}), and interchanging the surface and time integrals, yield
%
\begin{align}
\Delta \vv{L}^\text{mech}
=&
\oint_S
\frac{\epsilon_{kli}r_l}{4\pi}
\left\{
\int_{-\infty}^{\infty}
\left[
E_iE_j+B_iB_j
\right.
\right.
\noindent\nonumber\\
&
\left.
\left.
-
\frac{\delta_{ij}}{2}\left(E_\alpha E_\alpha +B_\alpha B_\alpha\right)
\right]
dt
\right\}
da_j.
\label{angularmomentum4}
\end{align}
%

It is convenient to recast Eq. (\ref{angularmomentum4}) in terms of a frequency integral instead of the time integral. For this purpose, let $A_j\left(\vv{r}, t\right)$ stand for $E_j$ or $B_j$. It is possible to express $A_j\left(\vv{r}, t\right)$ in terms of its time-to-frequency Fourier Transform $\tilde{A}_j\left(\vv{r}, \omega\right)=\fou{A_j\left(\vv{r}, t\right)}$, such that $A_j\left(\vv{r}, t\right) = \ifou{\tilde{A}_j\left(\vv{r}, \omega\right)}{\omega}$. Given that  $A_j\left(\vv{r}, t\right)$ is real, $\tilde{A}_i^{*}\left(\vv{r}, \omega\right)=\tilde{A}_i\left(\vv{r}, -\omega\right)$ (the symbol $*$ denotes complex conjugation), and Eq. (\ref{angularmomentum4}) is equivalent to
%
\begin{equation}
\Delta \vv{L}^\text{mech}
=
\int_{0}^{\infty}
\vec{\mathbfcal{L}}(\w)
d\w,
\label{TMA1}
\end{equation}
%
where $\vec{\mathbfcal{L}}(\w)$ is the \textit{spectral contribution to the angular momentum transfer}, whose $k$ component is given by
%
\begin{equation}
\mathcal{L}_k(\w)
=
\frac{1}{4\pi^2}
\oint_S
\epsilon_{kli}r_lD_{ij}\left(\vv{r},\w\right)
da_j,
\label{Lespectral}
\end{equation}
%
with
%
\begin{equation}
D_{ij}
=
\text{Re}
\left[
\tilde{E}_i\tilde{E}_j^{*} +\tilde{B}_i \tilde{B}_j^{*}
-
\frac{\delta_{ij}}{2}
\left(
\tilde{E}_\alpha \tilde{E}_\alpha^{*}
+
\tilde{B}_\alpha \tilde{B}_\alpha^{*}
\right)
\right],
\label{tensorD}
\end{equation}
%
in which $\tilde{E}_\alpha=\tilde{E}_\alpha\left(\vv{r}, \omega\right)$,  $\tilde{B}_\alpha=\tilde{B}_\alpha\left(\vv{r}, \omega\right)$, and $\text{Re}\left[z\right]$ denotes the real part of $z$.

The system is symmetric with respect to $xz$ plane, as can be seen from Fig. \ref{system}. Consequently, the forces and, therefore, the linear momentum that the swift electron transfers to the NP have no components in the $\hat{y}$ direction \cite{Lagos2, GarciadeAbajo-1,PRBCoronado}. In addition, the torque (about the center of the NP) and, thus, the angular momentum transferred to the NP cannot have components in $\hat{x} $ nor $\hat{z}$. That is, the NP rotates only about $y$ axis. Hence, using Eqs. (\ref{TMA1})-(\ref{tensorD}) and the aforementioned symmetry, it follows that
%
\begin{equation}
\Delta \vv{L}^\text{mech}
=
\left(\Delta L\right)  \hat{y}
=
\left(\Delta L_\text{E} + \Delta L_\text{M}\right)\hat{y},
\label{TMA4}
\end{equation}
%
where
%
\begin{align}
	\Delta L_\text{E}
	=&
	\frac{R^3}{4\pi^2}\!\!
	\int_0^\infty \!\!
	\int_0^{\pi}\!\!
	\int_0^{2\pi}\!\!\!
	\sin\theta\!
	\left(
	\cos\varphi 
	\text{Re}
	\left[
	\tilde{E}_\theta\tilde{E}_r^{*}
	\right]
	\right.
	\noindent\nonumber\\
	&
	\left.
	\!-\!
	\cos\theta \sin\varphi 
	\text{Re}
	\left[
	 \tilde{E}_\varphi \tilde{E}_r^{*}
	\right]
	\right)
	d\varphi d\theta d\w,
	\label{TMAE}
	\\
	\Delta L_\text{M}
	=&
	\frac{R^3}{4\pi^2}\!\!
	\int_0^\infty \!\!
	\int_0^{\pi}\!\!
	\int_0^{2\pi}\!\!\!
	\sin\theta\!
	\left(
	\cos\varphi 
	\text{Re}
	\left[
	\tilde{B}_\theta \tilde{B}_r^{*}
	\right]
	\right.
	\noindent\nonumber\\
	&
	\left.
	\!-\!
	\cos\theta \sin\varphi 
	\text{Re}
	\left[
	\tilde{B}_\varphi \tilde{B}_r^{*}
	\right]
	\right)
	d\varphi d\theta d\w,
	\label{TMAM}
\end{align}
%
are the electric and magnetic contributions to the angular momentum transfer, respectively, and spherical coordinates $r=\sqrt{x^2 + y^2 + z^2}$, $\theta=\arctan{\left(\sqrt{x^2 + y^2}/z\right)}$, and $\varphi=\arctan{\left(y/x\right)}$ have been used. Since the total electromagnetic fields are the sum of the external and scattered EM fields: $\vv{E}^\text{total}=\vv{E}^\text{ext}+\vv{E}^\text{scat}$ and $\vv{B}^\text{total} = \vv{B}^\text{ext}+\vv{B}^\text{scat}$, it is possible to express $\Delta L$ in Eq. (\ref{TMA4}) as
%
\begin{equation}
\Delta L
=
\Delta L^\text{ext-ext} + \Delta L^\text{scat-scat} + \Delta L^\text{ext-scat},
\label{TMA7}
\end{equation}
%
where
%
\begin{equation}
\Delta L^\text{a-b}
=
\Delta L_\text{E}^\text{a-b} + \Delta L_\text{M}^\text{a-b},
\label{TMAext-ext}
\end{equation}
%
in which the symbol ``a-b'' stands for ext-ext, scat-scat, or ext-scat. Explicitly, from Eqs. (\ref{TMAE}) and (\ref{TMAM}),
%
\begin{align}
\Delta L_\text{E}^\text{{ext-ext}}
=&
\frac{R^3}{4\pi^2}\!\!
\int_0^\infty \!\!
\int_0^{\pi}\!\!
\int_0^{2\pi}\!\!\!
\sin\theta\!
\left(
\cos\varphi 
\text{Re}
\left[
\tilde{E}_\theta^\text{ext} \tilde{E}_r^{\text{ext}*} 
\right]
\right.
\noindent\nonumber\\
&
\left.
\!-\!
\cos\theta \sin\varphi 
\text{Re}
\left[
\tilde{E}_\varphi^\text{ext} \tilde{E}_r^{\text{ext}*}
\right]
\right)
d\varphi d\theta d\w,
\label{TMAEext-ext}
\\
%
%
\Delta L_\text{E}^\text{scat-scat}
=&
\frac{R^3}{4\pi^2}\!\!
\int_0^\infty \!\!
\int_0^{\pi}\!\!
\int_0^{2\pi}\!\!\!
\sin\theta\!
\left(
\cos\varphi 
\text{Re}
\left[
\tilde{E}_\theta^\text{scat} \tilde{E}_r^{\text{scat}*} 
\right]
\right.
\noindent\nonumber\\
&
\left.
\!-\!
\cos\theta \sin\varphi 
\text{Re}
\left[
\tilde{E}_\varphi^\text{scat} \tilde{E}_r^{\text{scat}*}
\right]
\right)
d\varphi d\theta d\w,
\label{TMAEesp-esp}
\\
%
\Delta L_\text{E}^\text{ext-scat}
=
&\frac{R^3}{4\pi^2}\!\!
\int_0^\infty \!\!
\int_0^{\pi}\!\!
\int_0^{2\pi}\!\!\!
\sin\theta\!
\left(
\cos\varphi 
\text{Re}
\left[
\tilde{E}_\theta^\text{ext} \tilde{E}_r^{\text{scat}*}  
\right.
\right.
\noindent\nonumber\\
&
\left.
\left.
+
 \tilde{E}_\theta^\text{scat} \tilde{E}_r^{\text{ext}*} 
\right] 
-
\cos\theta \sin\varphi 
\text{Re}
\left[
\tilde{E}_\varphi^\text{ext} \tilde{E}_r^{\text{scat}*} 
\right.
\right.
\noindent\nonumber\\
&
\left.
\left.
+ 
\tilde{E}_\varphi^\text{scat} \tilde{E}_r^{\text{ext}*}
\right]
\right)
d\varphi d\theta d\w,
\label{TMAEext-esp}
\end{align}
%
and analogous formulas for the magnetic counterpart, obtained by replacing $\tilde{E}$ with $\tilde{B}$ in Eqs. (\ref{TMAEext-ext})-(\ref{TMAEext-esp}).

It is worth considering the case in which there is no nanoparticle, that is, the case of a free electron traveling in vacuum with constant velocity. In that case, nothing alters the electron's motion, so it loses no energy, momentum, nor angular momentum $\left(\Delta L = 0 \right)$. Applying the methodology presented here to this case, it follows from Eq. (\ref{TMA7}) that $\Delta L = \Delta L^\text{ext-ext} =0$, since there is no NP to scatter EM fields and a free electron loses no angular momentum. Hence, in general $\Delta L^\text{ext-ext}=0$ since it is physically equivalent to the angular momentum lost by a free electron. Therefore, 
%
\begin{align}
\Delta L_\text{E}
&=
\Delta L^\text{scat-scat}_\text{E} + \Delta L^\text{ext-scat}_\text{E},
\label{TMA7bis}
\\
\Delta L_\text{M}
&=
\Delta L^\text{scat-scat}_\text{M} + \Delta L^\text{ext-scat}_\text{M}.
\label{TMA7bisbis}
\end{align}
%
Then, the angular momentum transfer from the swift electron to a spherical nanoparticle, $\Delta\vec{L}^\text{mech}$, of any radius is finally obtained substituting Eqs. (\ref{TMA7bis}) and (\ref{TMA7bisbis}) into Eq. (\ref{TMA4}).

\subsection{Angular momentum transfer to small nanoparticles}

When the radius of a nanoparticle is much smaller than the impact parameter, the electromagnetic response of the NP is dominated by the dipole terms [$\ell=1$ in Eqs. (\ref{scatteredElectricField}) and (\ref{scatteredMagneticField})] \cite{FerrellEchenique1985,castrejon2021time}. This has led to studies for the linear and angular momentum transfers to small NPs modeled as point dipoles \cite{GarciadeAbajo, GarciadeAbajo-1, castrejon2021time, castellanos2021angular}, in what has been called \textit{small-particle approximation} (SPA). This approximation allows for more intuitive, easier, faster, and therefore more efficient momentum transfer calculations. 

Customarily, a particle is considered small relative to the wavelength of the external exciting EM field. However, in the present situation, it is not clear what does ``small'' mean because the swift electron produces an EM field containing all frequencies \cite{maciel2019electromagnetic,GarciadeAbajo-1}. Therefore, validity criteria must be established for the SPA applied to a given NP in terms of the relevant parameters of the problem: the impact parameter $b$, electron's speed $v$, and radius $a$ of the NP. For example, in the case of linear momentum transfer, it was found that for aluminum and gold NPs of $a=1$ nm, and for electrons traveling faster than $50\%$ the speed of light, the SPA corresponds to considering only $\ell=1$ in Eqs. (\ref{scatteredElectricField}) and (\ref{scatteredMagneticField}), and $b/a \geq 3$ \cite{castrejon2021time}.

In Ref. \cite{castellanos2021angular}, an expression for the angular momentum transfer to a small spherical NP in the SPA was obtained. The SPA expression of $\Delta\vec{L}^\text{mech}$ (originally in SI units in Ref. \cite{castellanos2021angular}) in Gaussian atomic units is
%
\begin{align}
\Delta\vec{L}^\text{mech}_\text{SPA}
=
-\hat{y}
\int_{0}^{\infty}
&
\frac{8\omega\lvert\omega\rvert}{\pi v^4 \gamma^3}
K_0\left(\frac{\lvert\omega\rvert b}{\gamma v}\right)
K_1\left(\frac{\lvert\omega\rvert b}{\gamma v}\right)
\noindent\nonumber\\
&
\times
\text{Im}
\left[
\frac{3i c^3}{2 \omega^3}  a_1\left(\omega\right)
\right]
d\omega,
\label{SPA}
\end{align}
%
where $K_0$ and  $K_1$ are the modified Bessel functions of order zero and one \cite{Abramowitz}, respectively, $\text{Im}[z]$ denotes the imaginary part of $z$, $a_1(\omega)$ is the first Mie coefficient \cite{Bohren}, and $\gamma=\left(1-\beta^2\right)^{-1/2}$ is the Lorentz factor, with $\beta=v/c$ and $c$ the speed of light.

In Subsection \ref{SPAresults}, we establish validity criteria for the SPA in the angular momentum transfer to aluminum, gold, and bismuth NPs of $a=1$ nm by comparing the results obtained from Eqs. (\ref{TMA4}) and (\ref{SPA}).

\subsection{Numerical considerations}
\label{Numericalconsiderations} 

Due to the short interaction time between a swift electron and a nanoparticle (on the order of tens of attoseconds), the causality of the dielectric function $\varepsilon(\omega)$ that characterizes the electromagnetic response of the NP is of paramount importance \cite{coronado2018, PRBcastrejon2}. For example, a non-causality on the order of the interaction time can change the sign of the linear momentum transfer  \cite{PRBcastrejon2}. Special care must be taken to compute the $\left(0, \infty\right)$ frequency integrals [see Eqs. (\ref{TMA1}) and (\ref{SPA})] because they always involve $\varepsilon(\omega)$. Typically, the dielectric function is only known over a frequency window, so interpolation and extrapolation are needed to compute the frequency integrals. Hence, to obtain reliable results for the angular momentum transfer, only causal $\varepsilon(\omega)$ must be used, and thus the causality of any inter- and extrapolation should be tested. Therefore, in this work we study the angular momentum transfer to NPs whose dielectric functions are knowingly causal, satisfying Kramers-Kronig relations \cite{Jackson}. As the first case of study, we consider aluminum NPs, whose electromagnetic response is characterized by Drude model with parameters given in Ref. \cite{Markovic} and then we consider gold and bismuth NPs, with more realistic dielectric functions taken from Werner \textit{et al.} work \cite{werner}, resulting from the fitting of experimental data to a Drude plus eight Lorentzian terms. 

Furthermore, given the complexity of the integrals involved in the computation of $\Delta L$, it is necessary to use numerical methods to calculate them. Recently, it has been shown that, in the problem of linear momentum transfer from a swift electron to a spherical nanoparticle, numerical convergence is crucial \cite{PRBcastrejon2}. In fact, in Ref. \cite{PRBcastrejon2} it is shown that results with incorrect physical behaviors have been previously published due to using conventional integration methods without ensuring numerical convergence. 

To calculate the integrals that appear in Eq. (\ref{TMA4}) [see Eqs. (\ref{TMAEext-ext})-(\ref{TMAEext-esp})] it is convenient to use the numerical methods known as \textit{cubatures} \cite{cools2002advances, hahn2007cuba, stroud1971approximate, berntsen1991adaptive}, which represent the state of the art in the calculation of multiple integrals and with which it is possible to have a precise control of the error that is committed when integrating. In this work, we employed the CUHRE adaptative cubature (\textit{CUbature Routine in Hyperrectangular REgions of integration}) from Ref. \cite{hahn2007cuba} to calculate $\Delta L$ from Eq. (\ref{TMA4}). Moreover, to compute the integral in Eq. (\ref{SPA}) we used the double-exponential Exp-Sinh quadrature \cite{mori2001double}. It is worth mentioning that the improper $(0, \infty)$ integrals are calculated ``exactly'' (with error control) by adequate change of variables that transform this interval into a bounded one \cite{kahaner1989numerical, mori2001double}. This eliminates the typical error by truncation arising from approximating $(0, \infty)$ by $(0, \omega_\text{cut})$, where $\omega_\text{cut}$ is an upper bound for the integration. In particular, all the results in this work have been calculated ensuring that their first three significant digits are correct.

The adaptative nature of CUHRE economizes the number of sampling points necessary to compute the integrals. This allows the accurate computation of $\Delta\vec{L}^\text{mech}$ for big NPs, with sizes covering the nanoscale (1-100 nm), as we show for aluminum NPs with radii up to 50 nm in Subsection \ref{FirstcaseofstudyAluminum}. It is worth mentioning that the same could be done for the linear momentum transfer. To the best of our knowledge, this is the first time that this type of calculation is performed on such big NPs, ensuring numerical convergence, without making further approximations.

%
%

\section{Angular momentum transferred to aluminum, gold, and bismuth nanospheres}

Using Eq. (\ref{TMA4}) and the numerical methodology discussed in Subsection \ref{Numericalconsiderations}, it is possible to compute the angular momentum transfer from a swift electron (of a STEM electron beam of any given energy) to a spherical NP of any material and diameter within the nanoscale (1-100 nm). 

In particular, in the small-particle approximation, the angular momentum transfer is directly related to the light-extinction properties of NPs \cite{castellanos2021angular}. Therefore, since plasmonic materials are prototypical absorbing materials, in this Section we present an implementation of our methodology for plasmonic nanospheres with knowingly causal dielectric functions. 


As the first case of study, for simplicity, we consider aluminum (Al) nanoparticles characterized by a Drude dielectric function with the following parameters: $\hbar\w_p=13.142$ eV and $\hbar\Gamma=0.197$ eV \cite{Markovic}. Next, for a more realistic case, we consider gold (a classic plasmonic material) and bismuth (a novel plasmonic material \cite{toudert2012exploring}) NPs, whose dielectric functions correspond to a fit of REELS data to a Drude plus eight Lorentzian terms reported in Werner \textit{et al.} work \cite{werner}. For all the materials studied, bulk dielectric functions are used because the size corrections are negligible for the angular momentum transfer calculations, as discussed in the \hyperlink{AppendixA}{Appendix}. Also, for a better understanding of the orders of magnitude, $\Delta L$ is expressed as a multiple of the reduced Planck's constant, $\hbar$, in all the figures of this work.

\subsection{First case of study: aluminum nanoparticles with Drude response}
\label{FirstcaseofstudyAluminum} 

To show the capabilities of both the theoretical and numerical methodologies developed in this work, in this Subsection we show a proof of concept using Al NPs with a Drude response. 

Since the integrals in Eqs. (\ref{TMAEext-ext})-(\ref{TMAEext-esp}) can be computed automatically with a prescribed precision for a given set of variables $\left\lbrace a, b, v,\varepsilon(\omega) \right\rbrace$  by the CUHRE algorithm, the only remaining parameter to be determined is the maximum number of multipoles $\ell_\text{max}$ to consider in Eqs. (\ref{scatteredElectricField}) and  (\ref{scatteredMagneticField}), instead of $\infty$. Therefore, for a given NP, it is necessary to find $\ell_\text{max}$ to achieve the desired precision. To illustrate this, we begin the study of Al NPs by presenting the multipolar contributions to the angular momentum transfer. Then, we show angular momentum results as a function of the impact parameter $b$ and the speed $\beta=v/c$ for Al NPs with radii covering the nanoscale, indicating in each case the $\ell_\text{max}$ required to ensure that the first three significant digits of the numerical results are correct.

\subsubsection{Multipolar contributions to the angular momentum transfer}

When the radius of a NP is much smaller than the impact parameter $\left( a\ll b\right)$, the electromagnetic response of the NP is mainly dipolar \cite{castrejon2021time, FerrellEchenique1985}. Hence, in this regime, taking $\ell_\text{max}=1$ is a good approximation. However, for arbitrary values of $b$ and $\beta$, more multipoles are necessary to ensure the accuracy in the calculation of $\Delta L$ in Eq. (\ref{TMA4}). The value of $\ell_\text{max}$ for a desired accuracy depends on the material and the radius of the NP. 

Figure \ref{Fig.multipolar-contributions}(a) shows $\Delta L$ as a function of $b$, for different values of $\ell_\text{max}$, in the case of an Al NP of radius $a=5$ nm interacting with a swift electron with speed $\beta=0.6$. Since the energy of the STEM electron beam corresponds to the kinetic energy of the swift electrons, $T=\left[\left(1-\beta^2\right)^{-1/2}-1\right]mc^2$ \cite{Jackson}, with rest mass $m$, the speed $\beta=0.6$ corresponds to an electron beam of $\approx 127.7$ keV. The marks in Fig. \ref{Fig.multipolar-contributions}(a) correspond to the calculated values of $\Delta L$, while the lines joining them are a guide to the eye (this will be the case for all the subsequent plots). The orange circles, indicating the results with $\ell_\text{max}=1$, correspond to the dipole contribution; the blue squares ($\ell_\text{max}=2$) are the combined contribution of the dipole plus quadrupole terms; $\ell_\text{max}=3$ corresponds to the sum of dipole, quadrupole, and octupole contributions; and so on.

%
\begin{figure}[h]
	\centering
	\includegraphics[width=\linewidth]{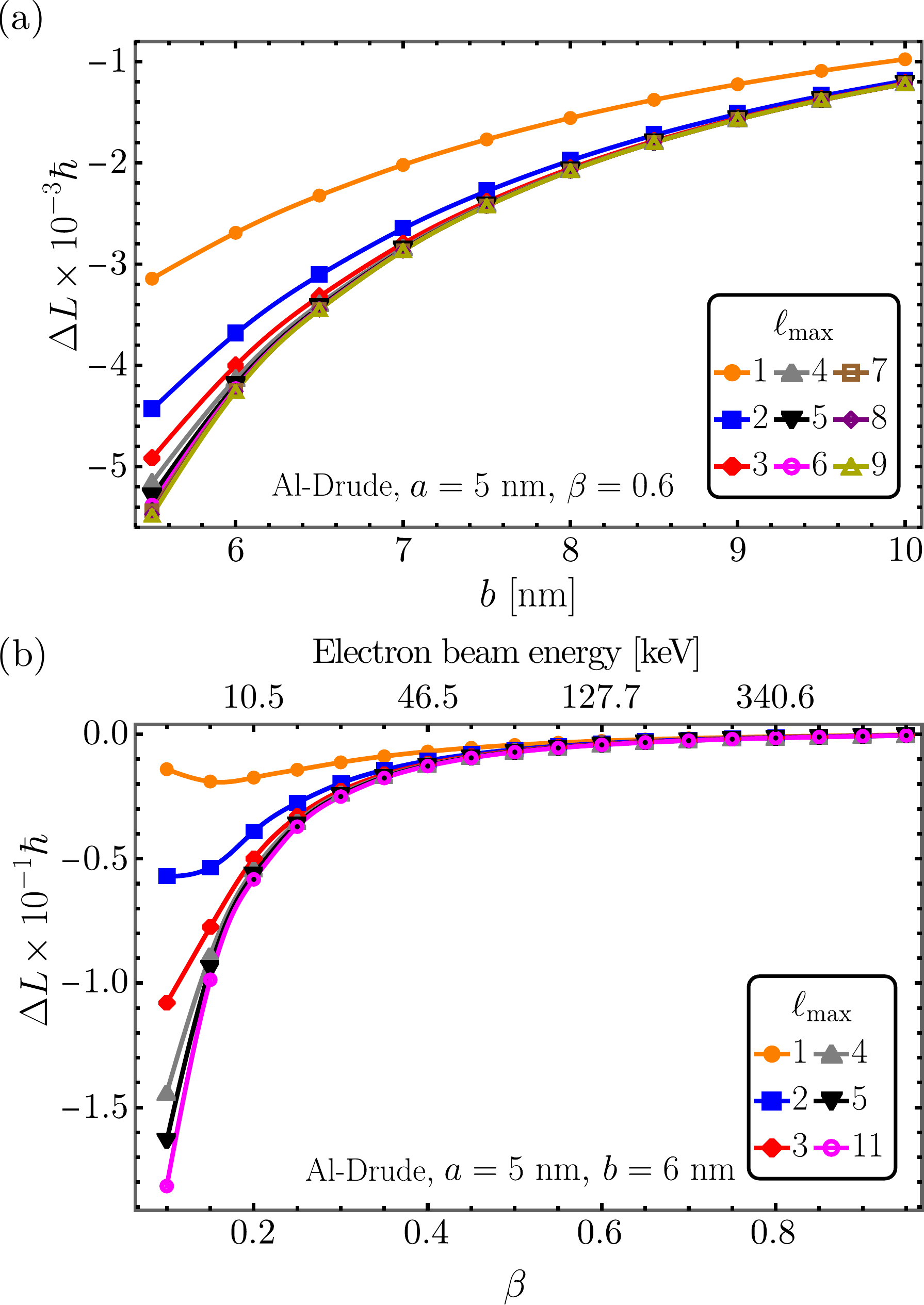}
	\caption{Multipolar contributions to the angular momentum transfer, $\Delta L$, from a swift electron to an aluminum nanoparticle, of radius $a=5$ nm, with Drude response \cite{Markovic} (a) as a function of the impact parameter, $b$, for a fixed electron's speed $\beta=v/c=0.6$, and (b) as a function of $\beta$ for a fixed $b=6$ nm. Solid lines are a guide to the eye.}
	\label{Fig.multipolar-contributions}
\end{figure}
%

It can be observed in Fig. \ref{Fig.multipolar-contributions}(a) that $\Delta L<0$ in all cases, indicating that the NP rotates around $-\hat{y}$ [see Fig. \ref{system} and Eq. (\ref{TMA4})]. Furthermore, all the multipolar terms contribute to the angular momentum transfer in such a way that there is a monotonic convergence of $\Delta L$ as a function of $\ell_\text{max}$. As expected, as $b$ increases, $\Delta L$ goes to 0, and fewer multipoles are needed to compute its value. In addition, for the NP under consideration, we found that the relative error of $\Delta L$ with $\ell_\text{max}=10$ (not shown in the figure) was less than $10^{-4}$. Therefore, to ensure that the first three significant digits of $\Delta L$ for an Al NP with $a=5$ nm are correct, it is sufficient to consider $\ell_\text{max}=10$ in Eqs. (\ref{scatteredElectricField}) and  (\ref{scatteredMagneticField}). 

Analogous observations can be made for $\Delta L$ as a function of $\beta$, whose plots are shown in Fig. \ref{Fig.multipolar-contributions}(b) for a fixed value of $b=6$ nm (so that the electron travels in a path 1 nm away from the NP's surface; see Fig. \ref{system}). The electron-beam-energy scale is shown on the upper horizontal axis. In particular, as expected, when $\beta\rightarrow 1$ fewer multipoles are needed and, in all cases,  $\Delta L$ approaches 0. 
%

Interestingly, for all the cases studied in this work (all materials, radii, and different combinations of parameters), the general characteristics discussed for Fig. \ref{Fig.multipolar-contributions} hold: (\textit{i}) there is a monotone convergence of $\Delta L$ as function of $\ell_\text{max}$, (\textit{ii}) $\Delta L<0$ always, indicating that the NP rotates around $-\hat{y}$, and (\textit{iii}) $\Delta L$ approaches 0 as $b$ or $\beta$ increases.

In the following, we present results for $\Delta L$ (for a given NP) ensuring that its first three significant digits are correct and indicating the value of $\ell_\text{max}$ needed to achieve this precision.

\subsubsection{Aluminum nanoparticles with sizes  covering the nanoscale range}

The methodology presented in Section \ref{Theoreticalmodel} allows  the calculation of angular momentum transfers to NPs with sizes covering the nanoscale. In Fig. \ref{Fig. Al-radii-nanoscale} we show results for $\Delta L$ to Al NPs of radii 5 nm [Figs. \ref{Fig. Al-radii-nanoscale}(a) and \ref{Fig. Al-radii-nanoscale}(b)], 10 nm [Figs. \ref{Fig. Al-radii-nanoscale}(c) and \ref{Fig. Al-radii-nanoscale}(d)], 20 nm [Figs. \ref{Fig. Al-radii-nanoscale}(e) and \ref{Fig. Al-radii-nanoscale}(f)] and 50 nm [Figs. \ref{Fig. Al-radii-nanoscale}(g) and \ref{Fig. Al-radii-nanoscale}(h)]. 
%
\begin{figure*}
	\includegraphics[width=1\textwidth]{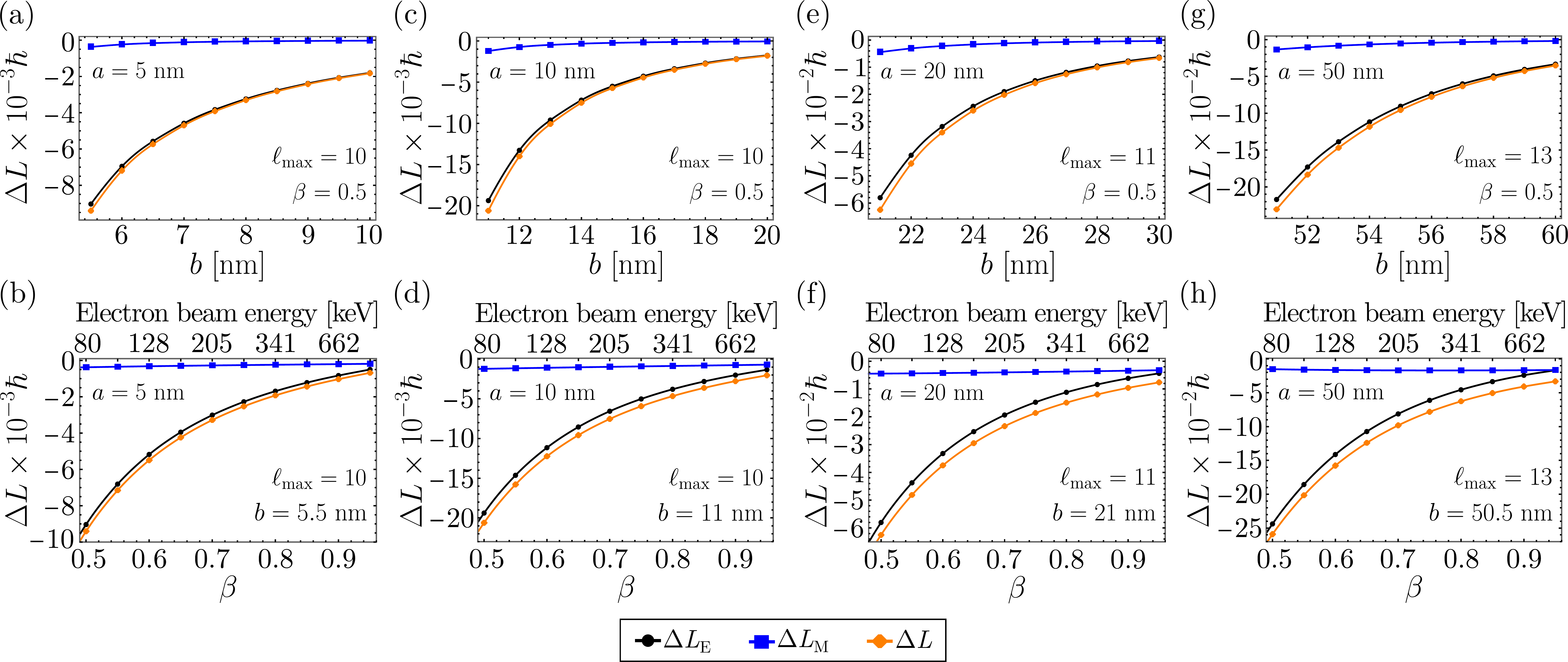}
	\caption{ 
		Angular momentum transfer ($\Delta L$, in orange diamonds) from a swift electron to aluminum nanoparticles of different radii covering the nanoscale, separating the electric $\left(\Delta L_\text{E} \text{, in black circles}\right)$ and magnetic $\left(\Delta L_\text{M} \text{, in blue squares}\right)$ contributions [see Eq. (\ref{TMA4})]. The value of the relevant parameters for the calculations are indicated in the insets, including the $\ell_\text{max}$ required to achieve the prescribed precision. (a) and (b) correspond to the results for NPs of radius $a=5$ nm, (c) and (d) for $a=10$ nm, (e) and (f) for $a=20$ nm, and (g) and (h) for $a=50$ nm. Solid lines are a guide to the eye.
	}
	\label{Fig. Al-radii-nanoscale}
\end{figure*}
%
%
In addition, following Eq. (\ref{TMA4}), we show the electric $\left(\Delta L_\text{E} \text{, in black circles}\right)$ and magnetic $\left(\Delta L_\text{M}\text{, in blue squares}\right)$ contributions to $\Delta L$ (shown in orange diamonds). In all cases, the value of $\ell_\text{max}$ necessary to achieve the prescribed precision is indicated in the inset, as well as the relevant parameters of the calculations.

In the top row of Fig. \ref{Fig. Al-radii-nanoscale}, we show $\Delta L$, $\Delta L_\text{E}$, and $\Delta L_\text{M}$ as a function of the impact parameter $b$ with a fixed value of $\beta=0.5$ (equivalent to a beam energy of 79 keV), while in the bottom row as a function of $\beta$ (with their beam-energy scales indicated) for fixed distances between the electron path and the NPs surfaces: $b-a=0.5$ nm for Figs. \ref{Fig. Al-radii-nanoscale}(b) and \ref{Fig. Al-radii-nanoscale}(h), and $b-a=1$ nm for Figs. \ref{Fig. Al-radii-nanoscale}(d) and \ref{Fig. Al-radii-nanoscale}(f). Since many STEM studies are performed with energies higher than 79 keV $\left(\beta=0.5\right)$ \cite{Krivanek1,Krivanek2,GarciadeAbajo-1,VERBEECK2005324,OLESHKO2013203}, we considered $\beta$ ranging from 0.5 to 0.95.

Notably, a general feature of Fig. \ref{Fig. Al-radii-nanoscale} is that the magnetic contribution to $\Delta L$ is practically negligible, so that $\Delta L$ is mainly due to the electric contribution. This is in agreement with the results obtained for small NPs in the SPA, for which $\Delta L$ has only electric contributions \cite{castellanos2021angular}. However, for high electron speeds, the magnetic contribution becomes more relevant for increasing radii (see bottom raw  of Fig. \ref{Fig. Al-radii-nanoscale}). Moreover, both $\Delta L_\text{E}$ and $\Delta L_\text{M}$ are, like $\Delta L$, always negative.

In all cases, $\Delta L$ approaches 0 as $b$ or $\beta$ increases. The same holds for $\Delta L_\text{E}$. However, $\Delta L_\text{M}$ approaches 0 just as $b$ increases, but moves away from 0 for high values of $\beta$ [this can be better appreciated in Fig. \ref{Fig. Al-radii-nanoscale}(h)].

As the radius of the NP increases, the angular momentum transferred from a swift electron also increases, from an order of magnitude of $10^{-3}\hbar$ (for $a=5$ nm) to $10^{-1}\hbar$ (for $a=50$ nm). Also, as expected, the value of $\ell_\text{max}$ needed to achieve the prescribed precision increases with the radius of the NP, being $\ell_\text{max}=10$ for $a=5$ nm, $\ell_\text{max}=10$ for $a=10$ nm, $\ell_\text{max}=11$ for $a=20$ nm, and $\ell_\text{max}=13$ for $a=50$ nm.

\subsection{More realistic materials: gold and bismuth nanoparticles}

When a dielectric function of more complex structure characterizes a plasmonic NP, as is the case for gold (Au) and bismuth (Bi) reported in Werner \textit{et al.} work \cite{werner}, the number of multipoles needed to achieve a given precision typically increases. Nevertheless, for these NPs, the main characteristics of the angular momentum transfer observed for the Drude Al case remain the same. We illustrate this for Au NPs in Figs. \ref{Fig.Au}(a) and \ref{Fig.Au}(b), and Bi NPs in Figs. \ref{Fig.Bi}(a) and \ref{Fig.Bi}(b), both with radius $a=5$ nm.

%
\begin{figure}[h]
	\centering
	\includegraphics[width=\linewidth]{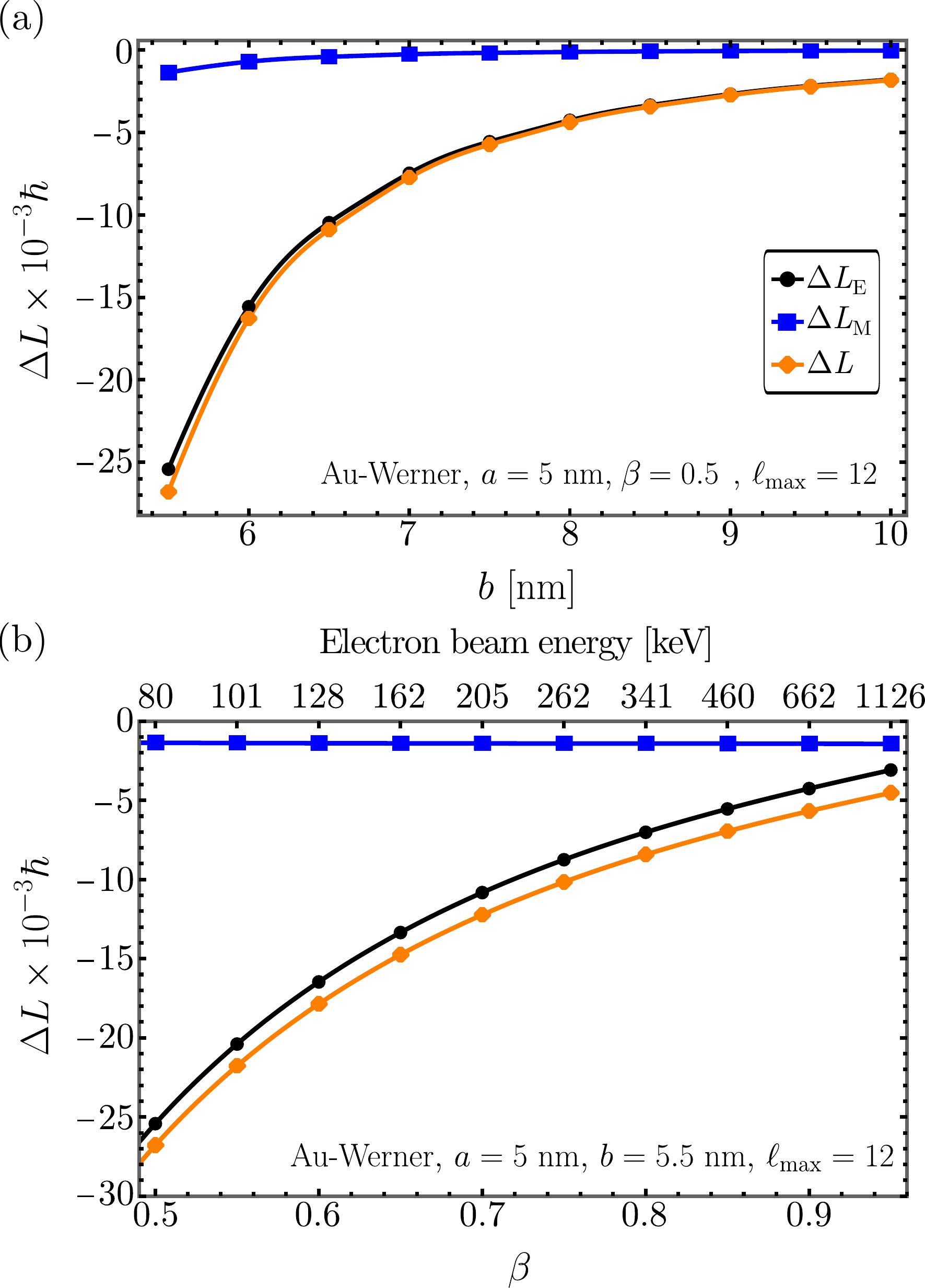}
	\caption{
Angular momentum transfer ($\Delta L$, in orange diamonds) from a swift electron to gold nanoparticles of radius $a=5$ nm (with dielectric function from Werner \textit{et al.} work \cite{werner}), separating the electric $\left(\Delta L_\text{E} \text{, in black circles}\right)$ and magnetic $\left(\Delta L_\text{M} \text{, in blue squares}\right)$ contributions [see Eq. (\ref{TMA4})]. (a) Results as a function of $b$ for fixed $\beta=0.5$, and (b) as a function of $\beta$ for fixed $b=5.5$ nm. Solid lines are a guide to the eye.
}
	\label{Fig.Au}
\end{figure}
%
%
\begin{figure}[h]
	\centering
	\includegraphics[width=\linewidth]{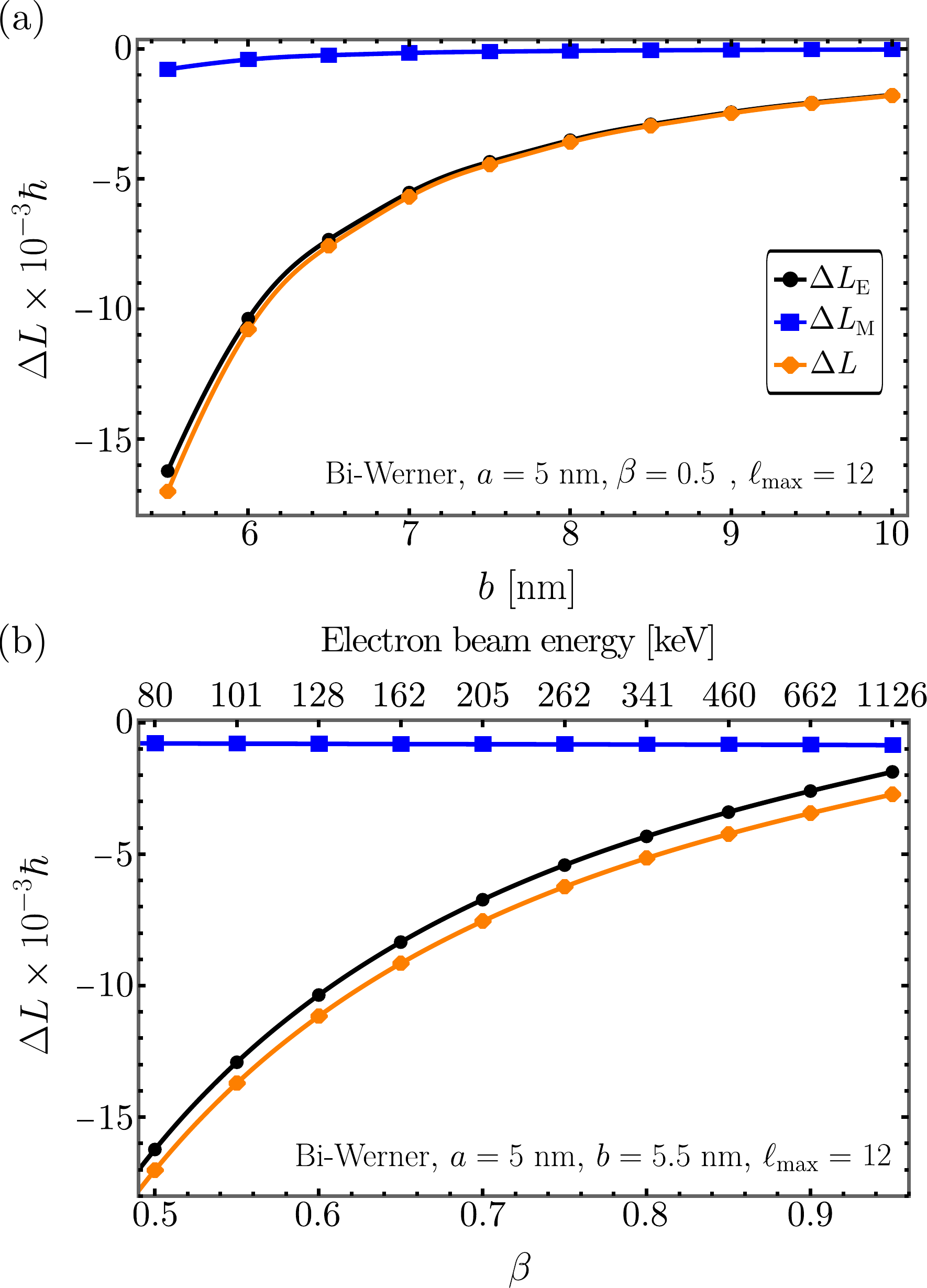}
	\caption{
Angular momentum transfer ($\Delta L$, in orange diamonds) from a swift electron to bismuth nanoparticles of radius $a=5$ nm (with dielectric function from Werner \textit{et al.} work \cite{werner}), separating the electric $\left(\Delta L_\text{E} \text{, in black circles}\right)$ and magnetic $\left(\Delta L_\text{M} \text{, in blue squares}\right)$ contributions [see Eq. (\ref{TMA4})]. (a) Results as a function of $b$ for fixed $\beta=0.5$, and (b) as a function of $\beta$ for fixed $b=5.5$ nm. Solid lines are a guide to the eye.
}
	\label{Fig.Bi}
\end{figure}

Figs. \ref{Fig.Au}(a) and \ref{Fig.Bi}(a) show $\Delta L$, $\Delta L_\text{E}$, and $\Delta L_\text{M}$ as functions of $b$ with a fixed $\beta=0.5$ (79 keV), while Figs. \ref{Fig.Au}(b) and \ref{Fig.Bi}(b) are the corresponding plots as a function of $\beta$ (and the electron beam energy) for fixed $b=5.5$ nm. In all cases, to achieve the prescribed accuracy of three correct significant digits, $\ell_\text{max}=12$ was necessary, in contrast to the Al NP of the same radius, for which $\ell_\text{max}=10$ was enough. 

The angular momentum transfer for Al, Au, and Bi NPs of $a=5$ nm is on the same order of magnitude, with the highest values occurring for gold and the lowest values for aluminum. Interestingly, the curves in Figs. \ref{Fig.Au} and \ref{Fig.Bi} are qualitatively equivalent to those presented in Fig. \ref{Fig. Al-radii-nanoscale}. The main differences are the specific values of $\Delta L$ and $\ell_\text{max}$. 

\subsection{Validity of the small-particle approximation}
\label{SPAresults}

As previously discussed, the number $\ell_\text{max}$ of multipoles necessary to achieve the desired accuracy decreases as $b$ and $\beta$ increase. However, $\ell_\text{max}$ also depends on the complexity of the dielectric function $\varepsilon(\omega)$. Interestingly, for small NPs, taking $\ell_\text{max}=1$ might be sufficient for large enough $\beta$ and $b/a$ ratios, depending on the NP's $\varepsilon(\omega)$. Moreover, in those conditions, the far simpler small-particle approximation (SPA) could ve valid, allowing for faster and more intuitive computations of the angular momentum transfer. Therefore, it is relevant to test the validity of the SPA for a given NP.
%
\begin{figure}[h]
	\centering
	\includegraphics[width=\linewidth]{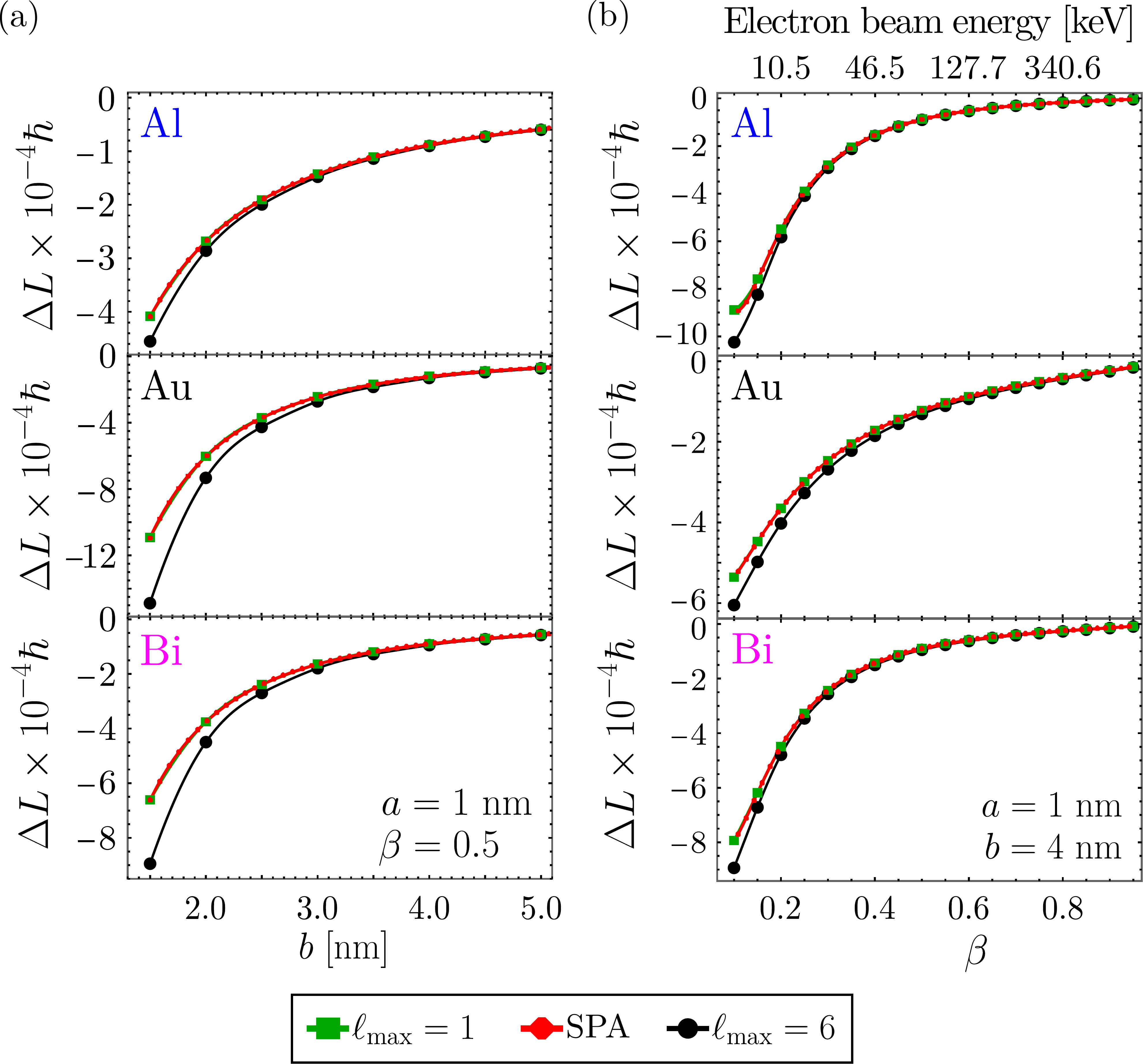}
	\caption{Angular momentum transferred to aluminum (top), gold (half), and bismuth (bottom) nanoparticles, of radius $a=1$ nm, as a function of (a) $b$ for fixed $\beta=0.5$, and (b) $\beta$ for fixed $b=4$ nm, calculated using $\ell_\text{max}=1$ (green squares), $\ell_\text{max}=6$ (black circles), and the small-particle approximation (SPA; red diamonds). Solid lines are a guide to the eye.
	 }
	\label{Fig.SPA-validation}
\end{figure}
%

In this Subsection, we establish validity criteria for the SPA in Al, Au, and Bi nanoparticles of radii $a=1$ nm. To do so, we compute $\Delta L$ from Eq. (\ref{TMA4}), using $\ell_\text{max}$ such that the relative error of the results is below $10^{-4}$, and compare the results with those obtained setting $\ell_\text{max}=1$, as well as with those of the SPA using Eq. (\ref{SPA}). 

First, we consider $\Delta L$ as function of $b$ for $\beta=0.5$ (79 keV); see Fig. \ref{Fig.SPA-validation}(a). Notably, the SPA results overlap with those of $\ell_\text{max}=1$. As expected, these calculations differ from the correct $\Delta L$ (computed with the prescribed accuracy using $\ell_\text{max}=6$) for small values of $b$. Interestingly, all results overlap for $b/a \agt 4$.

Then, to complement the test, we present in Fig. \ref{Fig.SPA-validation}(b) the corresponding $\Delta L$ as a function of $\beta$ (and the beam energy) for $b/a=4$, including also the range $\beta<0.5$. Interestingly, the calculations coincide for $\beta \agt 0.3$ in the case of Al and Bi nanoparticles, but it takes $\beta \agt 0.5$ for Au NPs. In addition, the calculations with $\ell_\text{max}=1$ and with the SPA overlap again. Therefore, SPA is equivalent to using $\ell_\text{max}=1$ for all the cases considered. This is particularly relevant in terms of computing time (SPA calculations were 2 orders of magnitude faster than the equivalent $\ell_\text{max}=1$).

Summarizing: the SPA for $\Delta L$ [Eq. (\ref{SPA})] in the considered NPs (\textit{i}) is equivalent to Eq. (\ref{TMA4}) taking  $\ell_\text{max}=1$ in Eqs. (\ref{scatteredElectricField}) and  (\ref{scatteredMagneticField}), and (\textit{ii}) is valid and accurate for $b/a \agt 4$ and $\beta \agt 0.5$.


\subsection{Comment on the orders of magnitude of the results}

Even though the angular momentum transferred to a nanoparticle by a single electron is very small, the number of incident electrons per unit of time in a STEM electron beam is very high. For example, in an electron beam with a current of 1 pA, there are $\sim 10^{7}$ electrons per second interacting with a nanoparticle. Therefore, there can be appreciable angular momentum transfers producing observable rotations of nanoparticles \cite{OLESHKO2013203}, as those reported in the experiments of Refs. \cite{Batson,Oleshko,xu2010transmission}. 

In all our results $\left\lvert \Delta L \right\rvert<\hbar/2$, which may appear as a contradiction with Quantum Mechanics. However, since we depart from a classical model, there is no real contradiction. Our results correspond to the classical limit of the interaction. Interestingly, an analogous observation for the linear momentum transfer was made in Ref. \cite{Oleshko}, concluding that the classical-limit calculations correspond to an average of a full quantum electrodynamics approach.

%
%

\section{Summary and conclusions}

We presented a classical electrodynamic theoretical model, along with an efficient numerical methodology, to calculate the angular momentum transferred from a swift electron, of a STEM electron beam, to a spherical nanoparticle. 
Using this methodology, we simulated the angular momentum transferred to aluminum, gold, and bismuth nanoparticles of different sizes.
In addition, we tested the applicability of the small-particle approximation (in which the nanoparticles are modeled as electric point dipoles) for the angular momentum transferred to nanoparticles with 1 nm radius of the studied materials.

The considered model system is symmetric with respect to the plane defined by the electron's trajectory and the line joining it with the center of the nanoparticle, corresponding to plane $xz$ in Fig. \ref{system}. As a consequence, the transferred angular momentum is always parallel to $y$ axis, perpendicular to this plane. Remarkably, in all our simulations, the transferred angular momentum was always in $-\hat{y}$ direction. This contrasts with the change of direction reported for analogous linear momentum transfer studies \cite{GarciadeAbajo0,PRBCoronado,Batson2,Lagos2,OLESHKO2013203}.

We found that, given the same conditions, gold presented the highest angular momentum transfers, while aluminum displayed the lowest ones. The order of magnitude of these transfers ranged from $10^{-4}\hbar$ (for aluminum, gold, and bismuth nanoparticles with radius 1 nm) up to $10^{-1}\hbar$ (for aluminum nanoparticles with radius 50 nm). Our calculations, which correspond to the classical limit of the interaction, indicate that there can be appreciable rotations of nanoparticles in STEM studies due to the amount of incident electrons in the beam (e.g. $\sim 10^7$ electrons per second for a beam of 1 pA), as those reported in the experiments of Refs. \cite{Batson,Oleshko,xu2010transmission}.

All our simulations displayed a monotonic convergence of the angular momentum transfer as a function of the number of multipoles considered (in the electromagnetic fields scattered by the nanoparticle). In addition, its magnitude increased with the radius of the nanoparticle, but decreased as the speed of the electron or the impact parameter increased. 
Moreover, the contribution of the electric field to the angular momentum transfer dominated over that of the magnetic field, becoming comparable only for high electron's speeds (greater than $90\%$ of the speed of light). 

Additionally, for nanoparticles with 1 nm radius, we found that the small-particle approximation for the angular momentum transfer is valid and accurate as long as the impact parameter is greater than four times the nanoparticle's radius, and that the electron's speed exceeds $50\%$ of the speed of light (equivalent to electron beams with energies greater than 79 keV). 

The present study contributes to reduce the gap towards successful development of electron tweezers, with potential implications in the progress of electron microscopy techniques such as electron vortices. Further developments might benefit from an broader analysis including more general nanostructures, made of different materials and geometries, as well as the inclusion of the effects of a substrate.

%
%

\begin{acknowledgments}
This work was supported by UNAM-PAPIIT project DGAPA IN107122. The main results were obtained while J. \'A. C-R. and J. C-F. were Ph. D. students from Programa de Doctorado en Ciencias Físicas, Universidad Nacional Autónoma de México (UNAM), receiving scholarships 481497 and 477516, respectively, from Consejo Nacional de Ciencia y Tecnología (CONACyT), Mexico. The authors are grateful to Prof. Rub\'en G. Barrera for fruitful discussions and valuable suggestions.
\end{acknowledgments}


\appendix*

%
%

\section{Size correction effects on the angular momentum transfer from a swift electron to a spherical nanoparticle}
\label{AppendixA}
\hypertarget{AppendixA}{}

There are different theoretical models for the dielectric function of a nanoparticle that depend on its dimensions \cite{kittel1996introduction}. In particular, it has been shown that a classical model can be used to characterize the electromagnetic response of spherical metallic nanoparticles with radii greater than 5 nm, while quantum effects are important for smaller nanoparticles \cite{kreibig2013optical,de2012quantum}. Interestingly, for metallic nanoparticles, different semiclassical and full-quantum studies arrive at the same conclusion: The width $\Gamma^*$ of the Mie resonances can be expressed as a sum of a size-independent contribution, $\Gamma_\text{bulk}$, and a size-dependent contribution, $\Gamma_\text{size}$ \cite{kreibig2013optical}. Moreover, for a spherical nanoparticle of radius $a$, many studies (both semiclassical and fully quantum) have found $\Gamma_\text{size}$ to be proportional to $a^{-1}$ (see references in Section 2.2 of Ref. \cite{kreibig2013optical}). Explicitly,
%
\begin{equation}
\Gamma_\text{size}(a)=A\frac{v_F}{a},
\label{sizecorrection}
\end{equation}
%
where $v_F$ is the Fermi velocity of the nanoparticle and $A$ is a constant (of the order of 1) whose specific value depends on the particular theoretical model used in the derivation of Eq. (\ref{sizecorrection}) \cite{kreibig2013optical,kraus1983plasmon}.

In our calculations we did not consider such size corrections because we found that, for the total angular momentum transfer, the corrections are negligible for the nanoparticle sizes we reported. To illustrate this fact, we show below the results for the smallest nanoparticle case reported in our work. 

In Fig. \ref{Fig.size-correction} we show the angular momentum transfer from a swift electron to an aluminum spherical nanoparticle of radius $a=1$ nm, using the Drude dielectric function employed in our work, with (blue line) and without (orange line) the size correction given by Eq. (\ref{sizecorrection}), as a function of the impact parameter $b$ [with electron's speed $\beta=v/c=0.5$, Fig. \ref{Fig.size-correction}(a)], and  the electron's speed $\beta$ [with impact parameter $b=1.5$ nm, Fig. \ref{Fig.size-correction}(b)].
%
\begin{figure}[h]
	\centering
	\includegraphics[width=\linewidth]{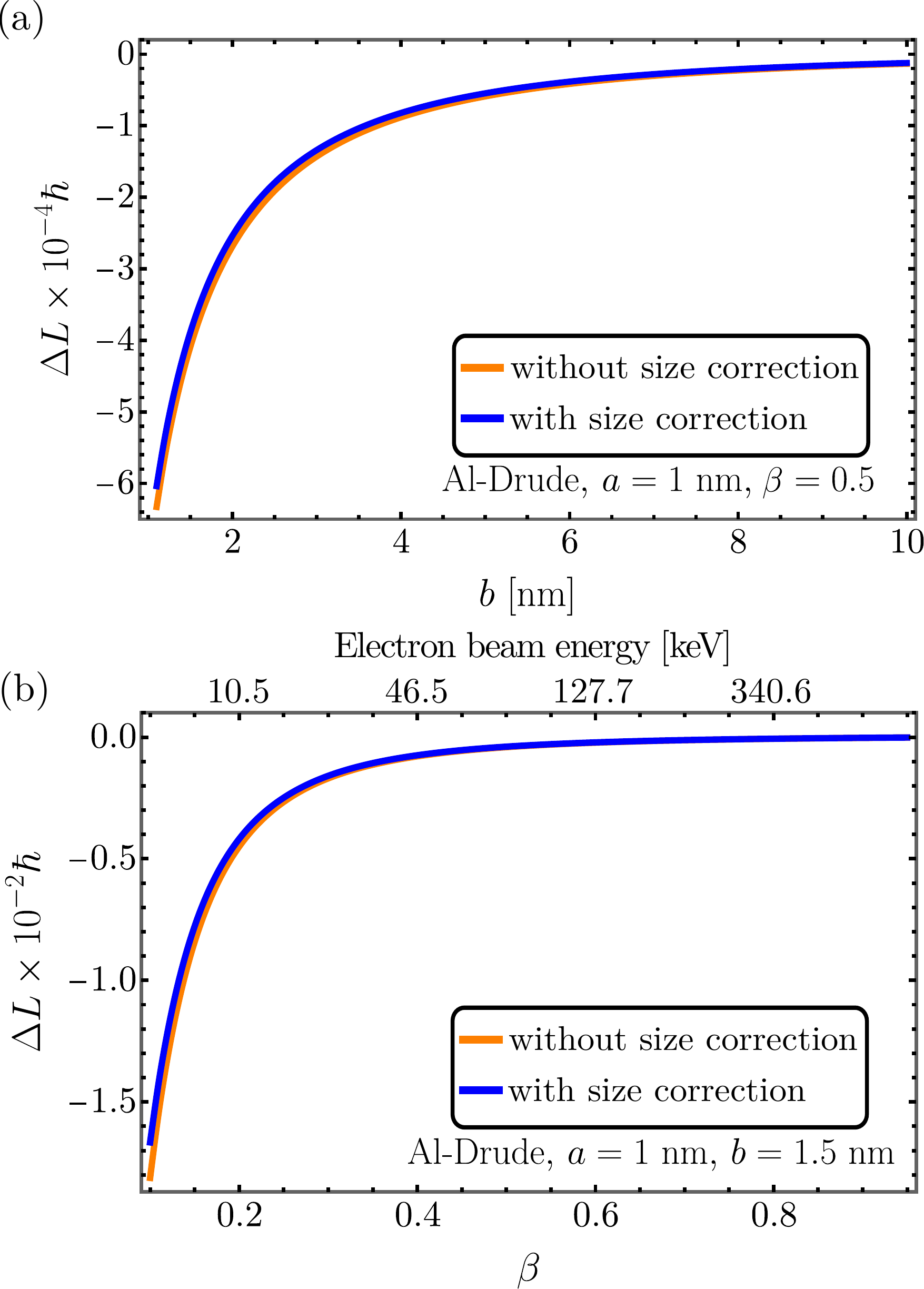}
	\caption{Angular momentum transfer from a swift electron to an aluminum nanoparticle of 1 nm radius (a) as a function of the impact parameter $b$, with a fixed electron's speed $\beta=v/c=0.5$, and (b) as a function of $\beta$ for fixed $b=1.5$ nm. The orange curve corresponds to the calculation using Drude bulk dielectric function $\varepsilon_\text{bulk}(\omega)$ [see Eq. (\ref{drudebulk})] without size correction, and the blue curve to the calculation with $\varepsilon_\text{corrected}(\omega, a=1\text{ nm})$ [see Eq. (\ref{drudecorrected})], including size correction.}
	\label{Fig.size-correction}
\end{figure}

Specifically, we used
%
\begin{equation}
\varepsilon_\text{bulk}(\omega)
=
1 - \frac{\omega_p^2}{\omega (\omega + i \Gamma_\text{bulk})}
\label{drudebulk}
\end{equation}
%
and
%
\begin{equation}
\varepsilon_\text{corrected}(\omega, a)
=
1 - \frac{\omega_p^2}{\omega [\omega + i (\Gamma_\text{bulk}+\Gamma_\text{size})]},
\label{drudecorrected}
\end{equation}
%
where $\hbar\omega_p=13.142\,\text{eV}$, $\hbar\Gamma_\text{bulk}=0.197\,\text{eV}$ \cite{Markovic} ($\hbar$ is Planck's reduced constant) and $\Gamma_\text{size}$ given in Eq. (\ref{sizecorrection}), using $A=1$ \cite{kreibig2013optical}, and the Fermi velocity for aluminum $v_F=2.03\times 10^6$ m/s \cite{Ashcroft}. 

As can be seen in Fig. \ref{sizecorrection}, the difference between the two models for the dielectric function is small even for the smallest particle considered ($a=1$ nm).

%
\bibliographystyle{apsrev4-2}
\bibliography{references}

\end{document}